# Planar chiral nanoantenna for excitation-chirality-controlled hot spot modulation and emitter-coupled circularly polarized emission

Abhik Chakraborty,[1,2,3,*] Xiaofei Wu,[3] Ankit Kumar Singh,[3] Fabian Scheidler,[4] Min Jiang,[2,3,5] Jürgen Popp,[1,2,3] Bert Hecht,[4] and Jer-Shing Huang[1,2,3,6,7,*]

[1]Institute of Physical Chemistry, Friedrich Schiller University Jena, Helmholtzweg 4, 07743 Jena, Germany
[2]Abbe Center of Photonics, Friedrich Schiller University Jena, Albert-Einstein-Str. 6, 07745 Jena, Germany
[3]Leibniz Institute of Photonic Technology, Albert-Einstein-Str. 9, 07745 Jena, Germany
[4]Experimental Physics 5, University of Würzburg, Am Hubland, 97074 Würzburg, Germany
[5]ARC Centre of Excellence for Transformative Meta-Optical Systems, Department of Electronic Materials Engineering, Research School of Physics, The Australian National University, Canberra, ACT 2600, Australia
[6]Research Center for Applied Sciences, Academia Sinica, 128 Sec. 2, Academia Road, Nankang District, Taipei 11529, Taiwan
[7]Department of Electrophysics, National Yang Ming Chiao Tung University, Hsinchu 30010, Taiwan

*Corresponding authors: abhik.chakraborty@uni-jena.de, jer-shing.huang@leibniz-ipht.de

**Abstract**

A planar chiral plasmonic nanoantenna exhibiting an excitation-chirality-dependent hot spot in a nanogap is numerically investigated. Additionally, the underlying design principles are examined, providing a broadly applicable framework for engineering chiral nanoantennas through controlled geometrical or modal asymmetry. The hot spot can be turned on and off by changing the handedness of the exciting circularly polarized light (CPL). This effect stems from the rationally designed interference of plasmonic modes excited by the linearly polarized orthogonal components of CPL. The hot spot exhibits maximal near-field dissymmetry factor ($\approx -2$) at a wavelength of 842 nm. The intensity at the hot spot can also be continuously modulated by varying the excitation ellipticity and handedness, approaching a modulation depth of 100%. These attributes enable chirality- and ellipticity-dependent switching and dynamic modulation of the plasmonic near field. Moreover, placing a quantum emitter in the gap generates almost perfectly circularly polarized emission, offering a simple yet effective avenue to realize nanoscale circularly polarized single-photon sources.

## 1. Introduction

General limitations to the efficiency of light−matter interaction at the mesoscopic to microscopic regime are the diffraction limit[1] and the size mismatch between the wavelength of light and an individual emitter.[2] Plasmonic nanostructures overcome these limitations and enhance weak light−matter interaction mainly through two complementary mechanisms. Firstly, plasmonic hot spots enforce the concentration of electromagnetic fields into deep sub-wavelength volumes, thus attaining sub-diffraction spatial resolution. This also leads to highly increased interaction probability through enhanced near-field intensity, enhanced local density of optical states, and a broadband near-field momentum distribution. Secondly, plasmonic nanostructures host electromagnetic modes that facilitate antenna-like radiative and resonator-like non-radiative behaviors, which can also be strategically engineered or selected to tailor coupling efficiencies and emission rates. In applications involving naturally forbidden light−matter interactions, such as intraband transitions typical of photoluminescence in gold[3] or brightening of dark excitons in 2D materials,[4] these complementary mechanisms enhance the coupling efficiency and radiative emission rate. In the case of nonlinear optics,[5,6] where phase-matching constraints, higher-order corrections to material polarizability, and small scattering cross-sections are crucial features, plasmonic nanostructures enhance the inherently weak interaction and resultant optical signals in a similar manner. The applicability of plasmonic nanostructures acting as rich playgrounds for light–matter interaction extends to domains as diverse as nanometrology,[7] photocatalysis,[8] color generation,[9] lasing,[10] quantum optics,[11,12] and more.

Coherent interaction or interference of spatially overlapping plasmonic modes in rationally designed nanostructures can be employed to optimally engineer vital attributes of the electromagnetic field, such as amplitude, phase, polarization, radiative directionality, pulse duration, group velocity, etc. One particularly compelling direction of research where interference of plasmonic modes becomes pivotal is the field of chiral nanoplasmonics.[13] Chirality is a geometric property characteristic of an object that cannot be superimposed on its mirror image. Chiral objects upon interaction with circularly polarized light (CPL) exhibit different response depending on the agreement or disagreement in handedness between the chiral object and CPL.[14] Much of the recent progress in chiral nanoplasmonics has centered on handedness-selective optical chirality enhancement in the near field for enantioselective sensing.[15] In contrast, this work theoretically demonstrates a planar chiral nanoantenna that shifts the focus from passive chiroptical sensing to the deterministic and excitation-chirality-controlled near-field switching for active manipulation of light−matter interaction. This is demonstrated by the excitation-chirality-dependent switchability of the nanoantenna's hot spot with continuous excitation-ellipticity-controlled tunability. Due to reciprocity, the nanoantenna can also effectively convert the radiative polarization of a dipolar quantum emitter into circularly polarized emission.

The first part of our work studies the on/off switching of a localized hot spot in a nanogap by changing the handedness of the illuminating CPL. The chiral sensitivity arises from the interference of plasmonic modes excited by the linearly polarized orthogonal components of CPL, allowing for the hot spot to be selectively activated or deactivated by simply switching the handedness of CPL. Notably, the nanoantenna demonstrates the value of the near-field dissymmetry factor (g) to be approximately $-2$ at the nanogap. It is important to note here that the maximum value of g is $\pm 2$. Moreover, by varying the excitation ellipticity and handedness, we show that the hot spot intensity can be continuously modulated with a modulation depth approaching 100%. The ability to turn on and off a hot spot by changing the handedness of CPL or to tune the near-field hot spot intensity as a function of optical ellipticity and handedness promises tunable near-field control of light−matter interaction, with applications like polarimetry,[16] polarization-sensitive circuitry,[17–19] light-driven nanomotors,[20,21] etc. In particular, emitter-coupled or nanostructure-induced coherent nonlinear processes[22–27] can act as instantaneous probes of the excitation-chirality-controlled hot spot. In the second part of this work, we demonstrate that when a dipolar quantum emitter is in the nanogap with the appropriate position and orientation, its emission polarization can possess a high degree of circular polarization (DoCP) over a large angular range. This allows for precise engineering of the handedness and polarization purity of emitted light, providing a deterministic route toward spin-polarized single-photon sources,[28,29] polarization-engineering plasmonic pixels,[30] and on-chip optical information processing.[31,32]

## 2. Geometry of the Planar Chiral Nanoantenna and Excitation-Chirality-Controlled Hot Spot

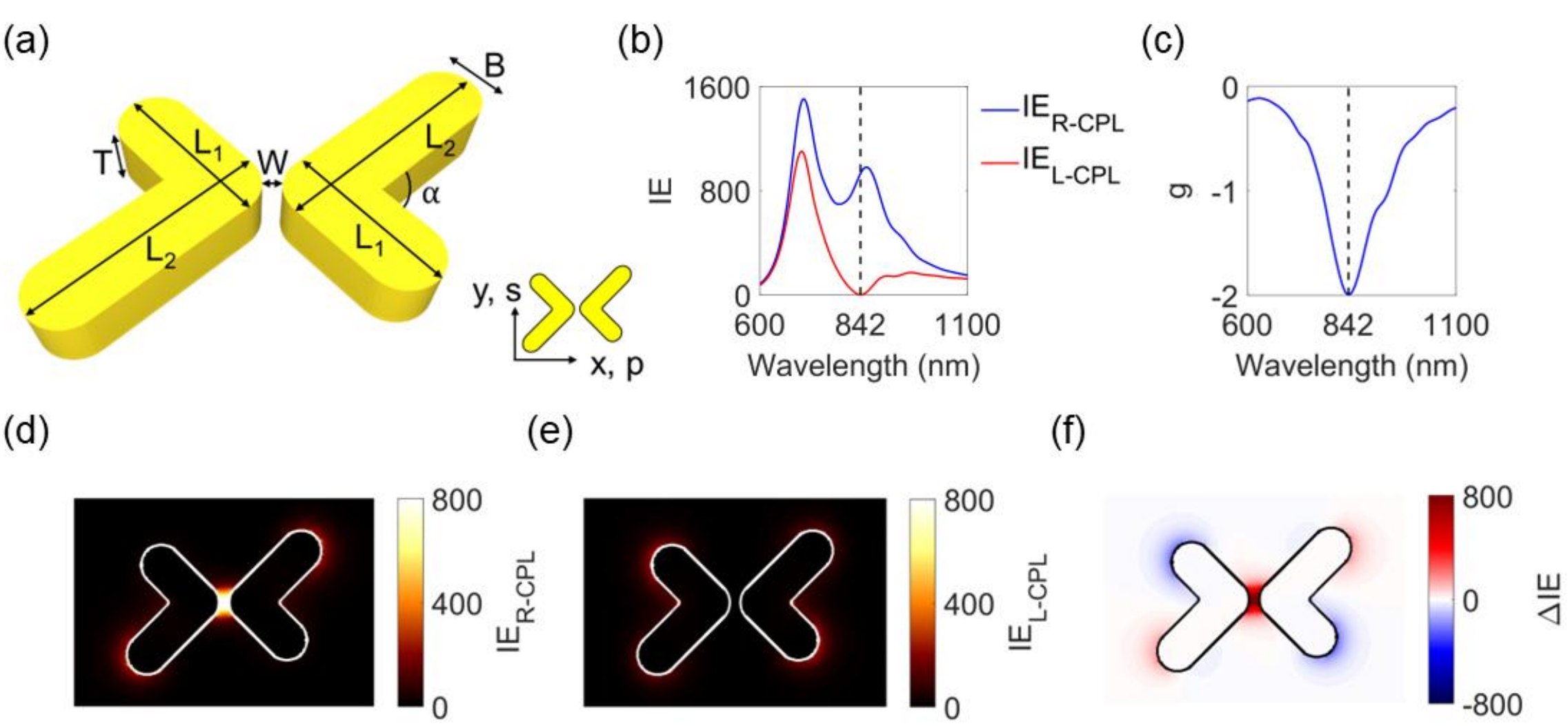

**Figure 1** (a) Scheme of a planar chiral nanoantenna consisting of two asymmetric V-shaped elements. Each element consists of two nanorods of unequal lengths which overlap at the apex of the V. The nanorods' lengths $L_1$, $L_2$, width B, and thickness T are 92 nm, 112 nm, 40 nm, and 30 nm, respectively. The angle between nanorods α and nanogap width W are 90° and 10 nm, respectively. The inset presents a top-view perspective of the nanoantenna. It elucidates that x and y are the in-plane near-field polarization states along and perpendicular to the nanogap axis, respectively, while p and s are the excitation polarization states along and perpendicular to the nanogap axis, respectively. (b) Intensity enhancement (IE) spectra at the center of the plasmonic nanogap under R-CPL and L-CPL excitation. The maximum contrast between $IE_{R\text{-}CPL}$ (blue) and $IE_{L\text{-}CPL}$ (red) is observed at a wavelength of 842 nm (vertical dashed line), where $IE_{R\text{-}CPL}$ is 931.68 and $IE_{L\text{-}CPL}$ is 0.98. (c) g spectrum exhibits maximal value of $\approx -2$ at 842 nm (vertical dashed line). (d, e) Spatial distribution of $IE_{R\text{-}CPL}$ and $IE_{L\text{-}CPL}$ at the wavelength of 842 nm calculated at a plane positioned mid-height of the nanoantenna. (f) Spatial distribution of ΔIE, i.e., difference between $IE_{R\text{-}CPL}$ and $IE_{L\text{-}CPL}$ at the wavelength of 842 nm calculated at the same plane. The planar chiral nanoantenna dimensions considered in the simulations are identical to the dimensions specified in Figure 1a.

The design criteria of the planar chiral nanoantenna demand two specific attributes: a single plasmonic hot spot and planar chiral symmetry breaking. The latter arises intrinsically in structures that lack mirror symmetry along both in-plane axes, ensuring distinct optical responses to differently handed CPL excitation. As depicted in **Figure 1a**, to achieve a single and well-defined hot spot, our nanoantenna consists of two V-shaped elements positioned with their apexes facing each other, forming a nanogap of width W. W is 10 nm. The two nanorods constituting each V-shaped element overlap at the aforementioned apex. This configuration ensures strong and localized field confinement at the nanogap, a prerequisite for achieving controlled near-field enhancement and modulation. However, simply using symmetric V-shaped elements would result in an achiral structure. To introduce chirality into the system, we start off by designing each V-shaped element with two coplanar nanorods of unequal lengths: a short rod of length $L_1$ and a long rod of length $L_2$. Therefore, each V-shaped element is intrinsically asymmetric. Furthermore, while both V-shaped elements use the same pair of nanorods with lengths $L_1$ and $L_2$, the relative positioning of the nanorods is inverted in one element in relation to the other. In the design central to this work and exhibited in Figure 1a, the left V-shaped element's nanorod with length $L_1$ extends into the upper-left quadrant and that with length $L_2$ into the lower-left quadrant from the top-view perspective. Conversely, for the right V-shaped element, the nanorod with length $L_2$ is at the upper-right quadrant and that with length $L_1$ is at the lower-right quadrant. The resultant geometry composed of two V-shaped elements collectively breaks mirror symmetry, and therefore, manifests planar chirality. In our design, $L_1$ is 92 nm and $L_2$ is 112 nm. Each nanorod has a width (B) of 40 nm and a thickness (T) of 30 nm. The angle ($\alpha$) between the two nanorods in each asymmetric V-shaped element is 90°. The chosen geometrical parameters satisfy the demonstrated fabrication capabilities of state-of-the-art ion beam milling techniques.[20,21,24,27] The inset of Figure 1a exhibits the top-view perspective of the nanoantenna, where x and y are the in-plane polarization states in the near field oriented along and perpendicular to the nanogap axis, respectively. Similarly, p and s are the excitation's linear polarization states oriented along and perpendicular to the nanogap axis, respectively. In this work, the handedness of CPL excitation is defined from the perspective of the receiver. L-CPL corresponds to a phase difference of $\phi_s - \phi_p = +90°$, while R-CPL corresponds to $\phi_s - \phi_p = -90°$.

Finite-Difference Time-Domain (FDTD) simulations are conducted to simulate the response of the nanoantenna depicted in Figure 1a situated on a glass substrate. To investigate the excitation-chirality-selective near-field response of the plasmonic nanoantenna, we simulate the intensity enhancement (IE) spectra at the geometric center of the nanogap under R-CPL and L-CPL illumination. The aforementioned IE spectra under R-CPL and L-CPL illumination are defined as $IE_{R\text{-}CPL}$ (blue in Figure 1b) and $IE_{L\text{-}CPL}$ (red in Figure 1b), respectively. A significant difference between $IE_{R\text{-}CPL}$ and $IE_{L\text{-}CPL}$ spectra is observed in Figure 1b, with the maximum contrast being observed at a wavelength of 842 nm, where $IE_{R\text{-}CPL}$ is 931.68 and $IE_{L\text{-}CPL}$ is 0.98. This disparity in near-field intensity enhancement is quantified using g as follows,

$$g = \frac{IE_{L-CPL} - IE_{R-CPL}}{IE_{L-CPL} + IE_{R-CPL}} \times 2 \tag{1}$$

Figure 1c shows that g at 842 nm is approximately $-2$, which is the maximum possible value. Spatial IE maps at 842 nm calculated in a plane situated at mid-height of the nanoantenna reveal a strong R-CPL-induced hot spot with $IE_{R\text{-}CPL}$ exceeding 800 within the nanogap (Figure 1d), in contrast to negligible $IE_{L\text{-}CPL}$ (Figure 1e). This is further confirmed in Figure 1f by the differential intensity enhancement map $\Delta IE$ ($IE_{L\text{-}CPL} - IE_{R\text{-}CPL}$) calculated at the same plane and wavelength. It can be seen that $\Delta IE$ is confined to the nanogap with its magnitude exceeding 800. This means that the designed nanoantenna is not only capable of yielding a single hot spot with maximal relative contrast in intensity ($g \approx -2$) but also provides a very high absolute contrast ($\Delta IE > 800$). These results collectively highlight both pronounced IE at the nanogap and a switchable single plasmonic hot spot with very high on/off ratio controlled by the handedness of CPL.

### 3. Origin of the Near-Field Chiroptical Response

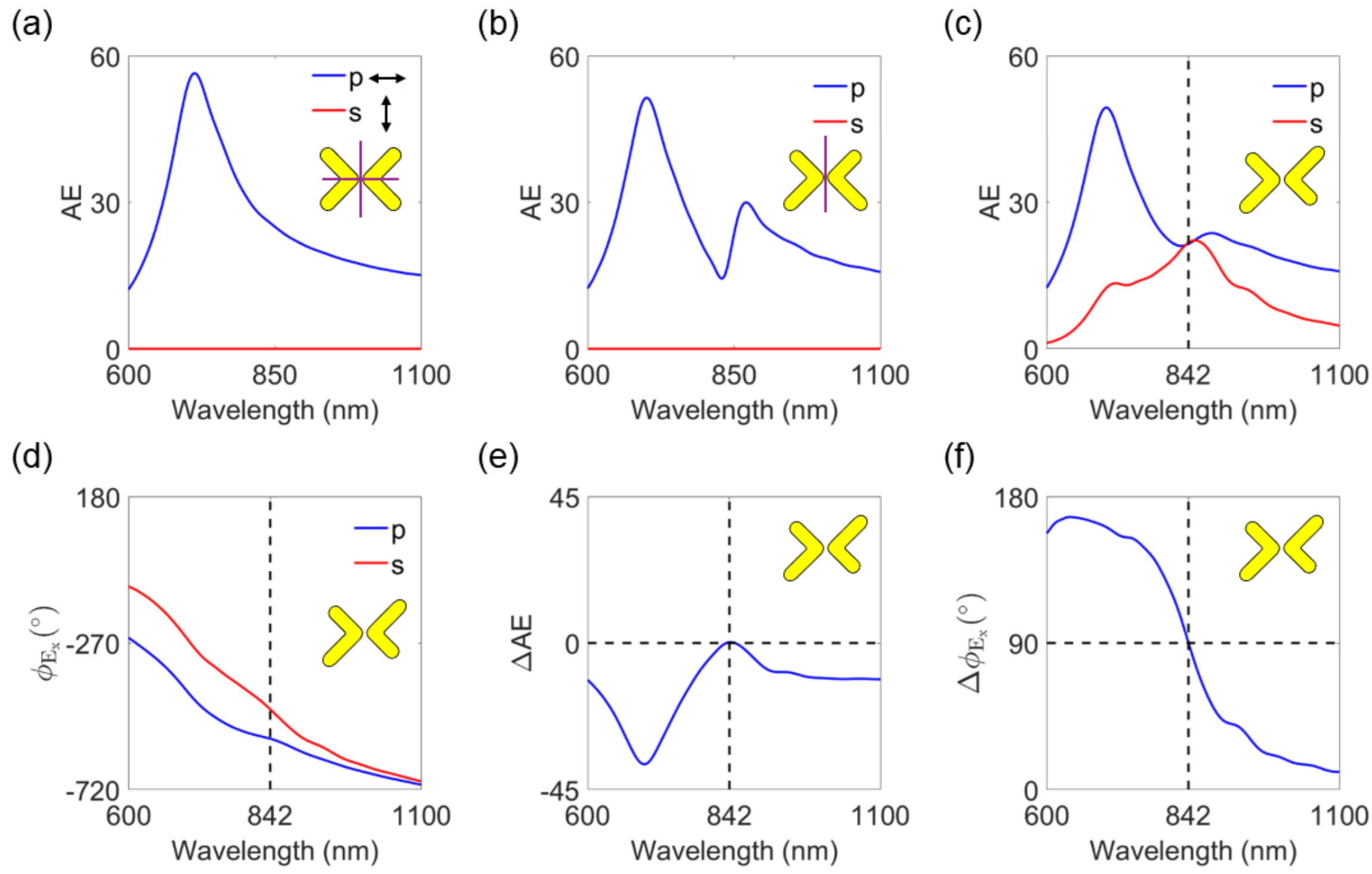

**Figure 2** Simulated amplitude enhancement (AE) of the x-polarized electric near field with respect to wavelength recorded at the center of the nanogap of (a) a fully symmetric, (b) a partially symmetric, and (c) a planar chiral double-V nanoantenna illuminated with p- (blue) and s-polarized (red) excitation. The insets show the nanoantenna geometries in all three cases, where the lines indicate the mirror planes in (a) and (b). (a, b) x-polarized AE spectra in the gaps of the fully symmetric and partially symmetric nanoantennas under p-polarized illumination, but no such response is observed under s-polarized illumination. (c) x-polarized AE spectra in the gap of the planar chiral nanoantenna under both p- and s-polarized illumination. (d) The phase spectra ($\phi_{E_x}$) of the co-polarized electric near field recorded at the gap of the planar chiral nanoantenna under p- and s-polarized illumination. At 842 nm (vertical dashed line), the co-polarized electric near fields co-localized in the nanogap exhibit (e) amplitude difference ΔAE approaching 0 (horizontal dashed line) and (f) phase difference $\Delta\phi_{E_x}$ around 90° (horizontal dashed line).

To achieve excitation-chirality-dependent switching of the plasmonic hot spot in the nanogap, the nanoantenna design must satisfy the following conditions:

i. Co-localization of the electric near field: The linearly polarized constituents of CPL, i.e., p- and s-polarized components, must excite plasmonic modes that produce spatially overlapping electric near fields at the nanogap.

ii. Co-polarization of the electric near field: Despite the orthogonality of the excitation polarization states, the co-localized electric near fields must have the same polarization.

iii. Equal amplitude contribution: The amplitudes of the co-localized and co-polarized electric near fields generated by p- and s-polarized excitations must be approximately equal at the nanogap.

iv. Phase difference of $\pm(2m+1) \times 90°$, $m \in \mathbb{N}_0$: The co-localized and co-polarized electric near fields of approximately equal amplitude generated by s- and p-polarized excitation must maintain a phase difference of approximately an odd multiple of 90°, i.e., $\pm(2m+1) \times 90°$, $m \in \mathbb{N}_0$. When driven under CPL excitation, an additional phase difference of ±90° is contributed by the handedness-dependent phase difference between the s- and p-polarized components of CPL. Therefore, the net phase difference can be written as $\pm\big((2m+1)\times 90° \pm 90°\big)$, $m \in \mathbb{N}_0$. Depending on the interplay of two factors, namely the parity of m and the agreement or disagreement between the signs of the modal phase difference at the nanogap and the excitation-chirality-dependent phase difference of CPL, the net phase difference equals either an even multiple of 180° ($\pm 2n \times 180°, n \in \mathbb{N}_0$) or an odd multiple of 180° ($\pm(2n+1) \times 180°, n \in \mathbb{N}_0$).

Therefore, these conditions collectively enforce a binary switching behavior of either constructive interference (hot spot “on”) with a net phase difference of an even multiple of 180°, or destructive interference (hot spot “off”) with a net phase difference of an odd multiple of 180°.[16] We compare three nanoantenna geometries considering the design criteria discussed above: fully symmetric geometry, partially symmetric geometry, and planar chiral geometry. The following section delves into each nanoantenna geometry to illustrate the role of structural asymmetry and consequent modal interference (see **Figure S1** in Supporting Information for additional information) in endowing chiroptical attributes in the optimal design.

Previously, it has been mentioned that the designed nanoantenna with planar chirality has two asymmetric V-shaped elements where each element has two rods of length $L_1$ (92 nm) and $L_2$ (112 nm) with inverted orientational configuration in relation to each other. In contrast to that, the fully symmetric nanoantenna shown in the inset of **Figure 2a** has two symmetric V-shaped elements, where all the rods have the identical length of $\frac{L_1+L_2}{2}$ (102 nm). Therefore, the structure has complete in-plane mirror symmetry. The mirror planes are indicated by the two orthogonally oriented lines in the inset. On the other hand, the partially symmetric nanoantenna depicted in the inset of Figure 2b consists of two asymmetric V-shaped elements, each with rod lengths $L_1$ and $L_2$. However, unlike the planar chiral design, the two V-shaped elements are not inverted relative to each other. This implies that the structure preserves mirror symmetry along one in-plane axis. The vertical line in the inset of Figure 2b is the axis across which mirror symmetry is preserved for the partially symmetric nanoantenna. Figure 2a and Figure 2b make it clear that p-polarized excitation (blue) generates x-polarized gap-confined amplitude enhancement (AE) in both cases. Both hot spots are predominantly x-polarized due to the geometry-imposed confinement along the same axis. However, Figure 2a and Figure 2b also show that s-polarized excitation does not generate x-polarized AE in the nanogap due to symmetry-induced restrictions in both cases (see Figure S1 in Supporting Information for additional information). Therefore, the design criteria are not met which, in turn, makes both geometrical configurations unsuitable for excitation-chirality-dependent near-field response.

The planar chiral configuration depicted in the inset of Figure 2c lacks mirror symmetry across both in-plane axes. As shown in Figure 2c, the in-plane asymmetry of the planar chiral nanoantenna ensures that both p- and s-polarized excitations generate co-polarized (x-polarized) hot spots co-localized in the nanogap. Very importantly, Figure 2c shows that the x-polarized fundamental mode recorded at the center of the nanogap under p-polarized and s-polarized excitation have approximately equal value at and around the wavelength of 842 nm. Since the primary design criteria of co-localization, co-polarization, and equal amplitude contribution have been satisfied for the fundamental mode of the planar chiral nanoantenna, the next step is to test the criterion for inducing a phase difference of $\pm(2m+1) \times 90°$, $m \in \mathbb{N}_0$ between the fundamental mode excited by s-polarized source and the same excited by p-polarized source at 842 nm. Figure 2d exhibits the phase $\phi_{E_x}$ (°) of the co-localized and co-polarized fields excited by the orthogonally polarized sources recorded at the center of the nanogap.

Figure 2e confirms that at the wavelength of 842 nm, the difference in amplitude enhancement ($\Delta\text{AE} = \text{AE}_\text{s} - \text{AE}_\text{p}$) of the two co-polarized fundamental modes co-localized at the nanogap under s-polarized and p-polarized excitation is approximately 0. Furthermore, Figure 2f shows that at 842 nm, the phase difference $\left(\Delta\phi_{\text{E}_\text{x}} = \phi_{\text{E}_{\text{x}_\text{s}}} - \phi_{\text{E}_{\text{x}_\text{p}}}\right)$ between the two nanogap modes is approximately 90°. The additional phase difference of $\phi_\text{s} - \phi_\text{p} = -90°$ associated with R-CPL excitation leads to a net phase difference of approximately 0°, leading to constructive interference and hot spot enhancement in the nanogap. Under L-CPL excitation ($\phi_\text{s} - \phi_\text{p} = +90°$), the net phase difference is approximately 180°, implying destructive interference in the nanogap. The conclusions explain the results exhibited in Figure 1b. Therefore, from Figure 2e and Figure 2f, it becomes clear that the planar chiral nanoantenna, owing to its in-plane structural asymmetry, satisfies the complete set of design criteria, thus making sure that CPL of opposite handedness leads to constructive or destructive interference in the nanogap which, in turn, switches the hot spot on or off.

## 4. Modulation of Hot Spot Intensity and g with Excitation Ellipticity and Handedness

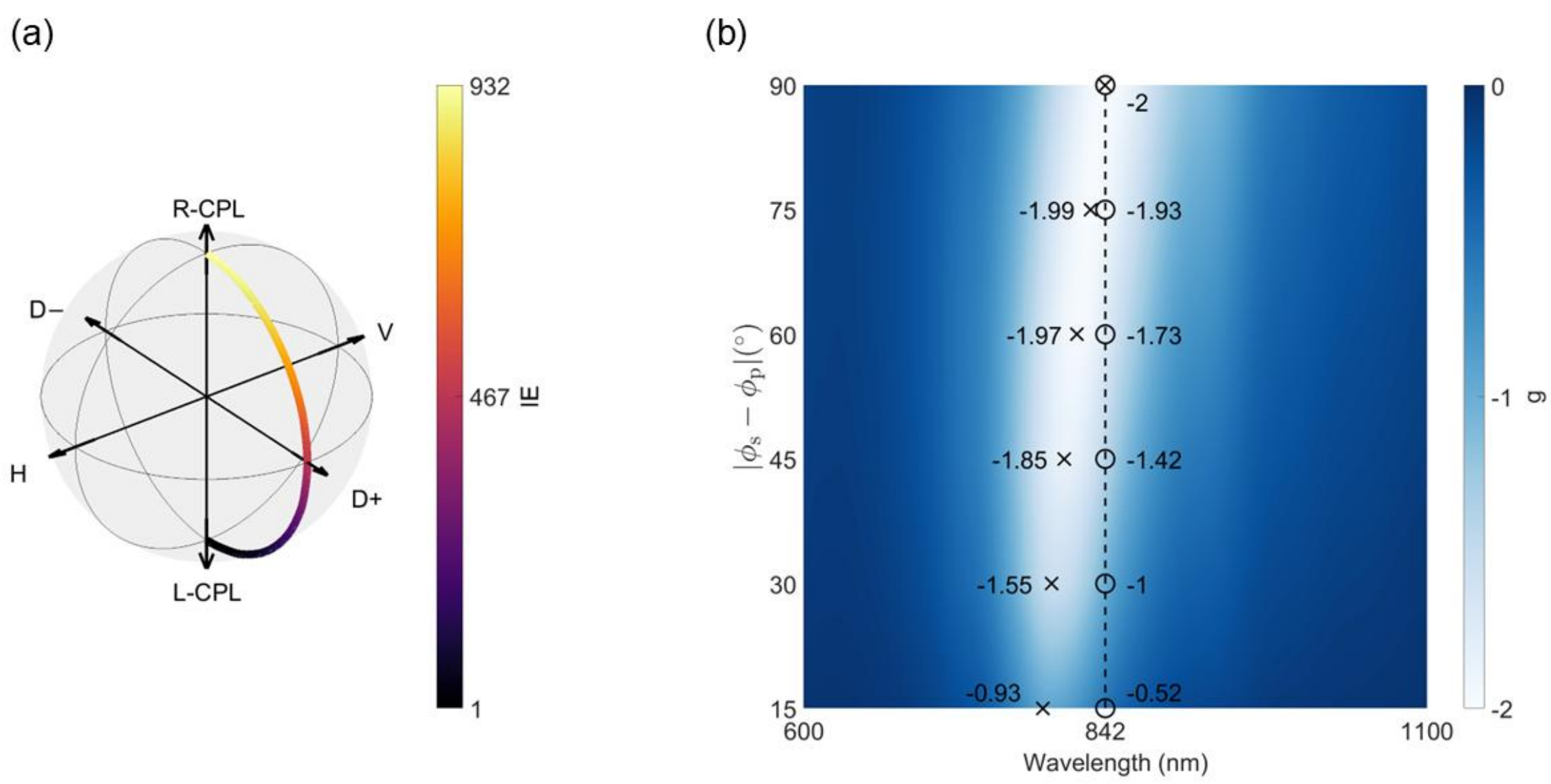

**Figure 3** (a) Poincaré sphere representation of IE at the nanogap at the wavelength of 842 nm showing continuous modulation with varying excitation polarization state. A trajectory from R-CPL to L-CPL via D+ polarization yields continuous and near-complete tunability, with IE peaking at 931.68 (approximated to 932) under R-CPL and decreasing monotonically to 0.98 (approximated to 1) under L-CPL. This corresponds to a modulation depth of approximately 100 %. (b) g as a function of excitation ellipticity, parameterized by the phase difference $|\phi_\text{s} - \phi_\text{p}|$. Maximal overall g of approximately $-2$ is seen at 842 nm under CPL $(|\phi_\text{s} - \phi_\text{p}| = 90°)$, while reduced phase difference lowers g (denoted by ×) monotonically to $-0.93$ at 798 nm for $|\phi_\text{s} - \phi_\text{p}| = 15°$. At 842 nm (vertically dashed line), g (denoted by ○) decreases from $-2$ (CPL) to $-0.52$ $(|\phi_\text{s} - \phi_\text{p}| = 15°)$.

**Figure 3** presents a comprehensive analysis of the chiral near-field response enabled by the planar chiral nanoantenna. In Figure 3a, a Poincaré sphere representation illustrates the modulation of IE at the center of the nanogap as a function of the excitation polarization state at the wavelength of 842 nm. A continuous trajectory from R-CPL to L-CPL incidence passing through the D+ linear polarization point (oriented at 45° to p) reveals a tunable response governed solely by the phase difference between the linearly polarized orthogonal components. IE reaches a value of 931.68 under R-CPL excitation and diminishes monotonically to 0.98 under L-CPL, demonstrating precise control of the hot spot intensity via incident polarization's handedness and ellipticity. The continuous tunability is remarkable as it achieves a modulation depth $\left(\left|\frac{I_{L-CPL}-I_{R-CPL}}{I_{R-CPL}}\right| \times 100\right)$ of approximately 100%. Figure 3b quantifies this behavior further by plotting g with respect to wavelength at various excitation ellipticities determined by the absolute phase difference between the linearly polarized orthogonal components of the excitation, i.e., $|\phi_s - \phi_p|$. The global maximum g of approximately $-2$ occurs at 842 nm under CPL $\left(|\phi_s - \phi_p| = 90°\right)$ while reducing phase difference or increasing ellipticity monotonically lowers g, reaching a local maximum of $-0.93$ at 798 nm for $|\phi_s - \phi_p| = 15°$. At the wavelength of 842 nm, one sees a similar trend in g ranging from $-2$ for CPL to $-0.52$ for elliptically polarized light with $|\phi_s - \phi_p| = 15°$.

## 5. Far-Field Response of Single Emitter-Driven Planar Chiral Nanoantenna

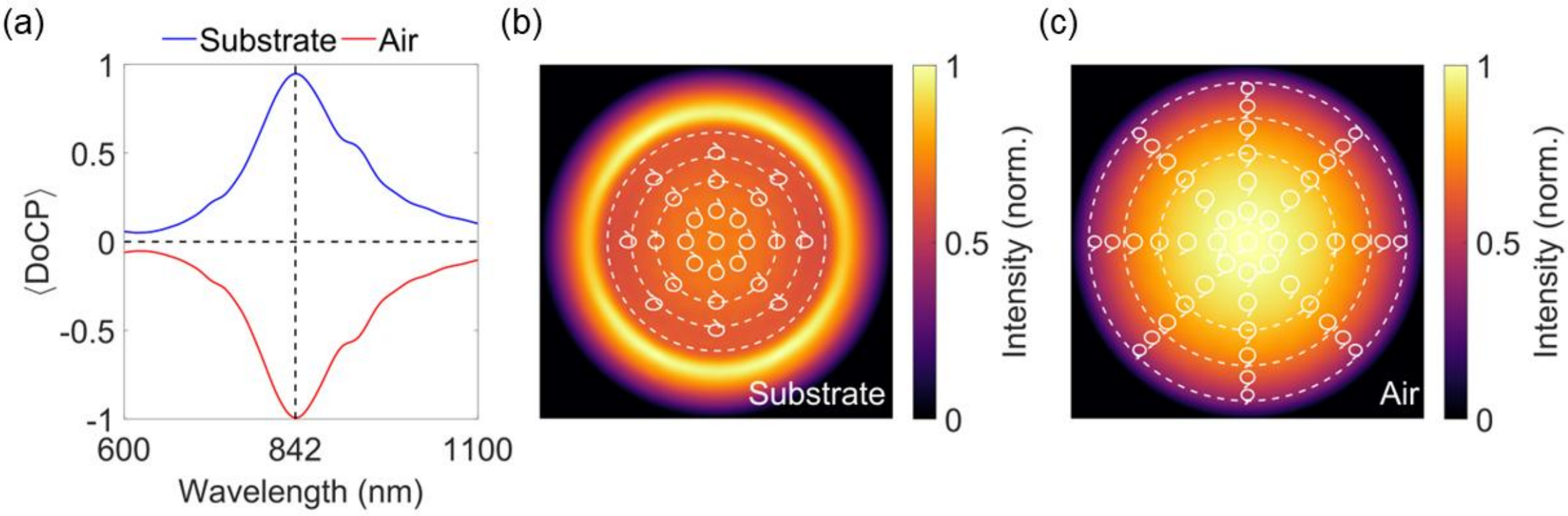

**Figure 4** (a) Average degree of circular polarization (⟨DoCP⟩) of far-field radiation within a numerical aperture (NA) of 0.9 from an x-polarized dipole emitter placed at the geometrical center of the planar chiral nanoantenna's nanogap. Across the spectrum, ⟨DoCP⟩ remains positive or left-handed (blue) on the substrate side and negative or right-handed (red) on the air side from the receiver's perspective. At 842 nm (vertical dashed line), ⟨DoCP⟩ of 0.86 and $-0.99$ are exhibited on the substrate and air sides, respectively. Normalized radiation patterns at 842 nm on the (b) substrate and (c) air sides with overlaid polarization ellipses. The three dashed circular rings in both plots exhibit the NA, where the innermost to outermost rings correspond to NA = 0.5, 0.7, and 0.9, respectively. The circularity of the polarization ellipses confirms robust and near-maximal DoCP over the entire NA of 0.9, while the handedness is confirmed to be left-handed on the substrate side and right-handed on the air side from the receiver's perspective.

Building on the demonstrated excitation-chirality-controlled hot spot switching, reciprocity dictates that the fully asymmetric nanoantenna should also impart spin selectivity to radiation from an emitter coupled to the hot spot. **Figure 4a** exhibits average DoCP (⟨DoCP⟩) of the far-field radiation from a single x-polarized dipole source located at the center of the nanogap. One can consider this dipole to be mimicking a single emitter in the gap. DoCP is the ratio of the Stokes parameters $S_3$ and $S_0$[33–35] and it can be mathematically described as follows,

$$\mathrm{DoCP}(\theta,\phi) = \frac{S_3(\theta,\phi)}{S_0(\theta,\phi)} = \frac{\Im\{E_\theta(\theta,\phi)E_\phi^*(\theta,\phi) - E_\phi(\theta,\phi)E_\theta^*(\theta,\phi)\}}{E_\theta(\theta,\phi)E_\theta^*(\theta,\phi) + E_\phi(\theta,\phi)E_\phi^*(\theta,\phi)} \tag{2}$$

$E_\theta$ and $E_\phi$ represent the complex electric fields in the polar and azimuthal directions in the far field, respectively. ⟨DoCP⟩ is calculated by dividing the integrated DoCP recorded in the far field across the numerical aperture (NA) of 0.9 by the surface area projected by the same NA's light cone. In Figure 4a, ⟨DoCP⟩ is positive or left-handed (blue) across the spectrum on the substrate side and negative or right-handed (red) on the air side from the perspective of the receiver, highlighting broadband spin-selective radiation enabled by the nanoantenna's planar chiral configuration. At 842 nm (vertical dashed line), the nanoantenna exhibits high ⟨DoCP⟩ of 0.86 on the substrate side and $-0.99$ on the air side over an NA of 0.9. Figure 4b and Figure 4c present the normalized radiated intensity pattern at 842 nm on the substrate and air sides, respectively. The polarization ellipses overlaid on the angular radiation patterns confirm that the polarization state is completely left-handed on the substrate side and right-handed on the air side across the entire NA of 0.9. As shown in **Figure S2** in Supporting Information, the radiated intensity is stronger into the substrate that into air. The contrast in intensity arises because the higher-index of substrate compared to that of air facilitates enhanced coupling to the propagating modes on the substrate side.[36]

## 6. Conclusion

In summary, we theoretically demonstrate a rationally designed planar chiral plasmonic nanoantenna that supports excitation-chirality-dependent modulation of hot spot and emitter-coupled far-field polarization engineering. The nanoantenna exhibits maximal g and modulation depth of the nanogap's hot spot intensity. This presents opportunities for diverse applications in nanoscale optoelectronics, polarimetry, nonlinear optics, etc. We also show that coupling a linearly polarized emitter to the nanogap yields far-field radiation with near-maximal ⟨DoCP⟩ over a large angular range, enabling deterministic generation of CPL. The high polarization purity and ⟨DoCP⟩ over an NA of 0.9, together with circularly polarized emission into both half-spaces, establish the planar chiral nanoantenna as a robust platform for quantum-emitter-driven, spin-resolved far-field applications, such as chiral single-photon sources. Furthermore, the interference of spectrally broad modes of the planar chiral nanoantenna is expected to be robust to moderate dimensional variations in fabricated nanostructures, with small mismatches preserving the polarization-selective nanoantenna response qualitatively. Importantly, the design principles discussed here are materially agnostic and rely primarily on modal interference with optimized relations in location, polarization, amplitude, and phase. Therefore, the design criteria are universal and can be employed to discover and optimize other nanoantenna geometries with tailored chiral responses.

## 7. Methods

### 7.1. Numerical Electrodynamic Simulations

Numerical simulations are conducted using the FDTD algorithm within a commercial 3D electromagnetic solver (Lumerical 2024 R1.1, Ansys). Total-field scattered-field plane-wave source at normal incidence is used to create the p-polarized, s-polarized, L-CPL, and R-CPL excitations. The nanoantennas are excited from the substrate side. For the far-field polarization analysis, an x-polarized dipole source is placed at the center of the nanoantenna's gap. The radiated fields from the dipole are recorded on the surface of a sphere of radius 1 m with the nanoantenna at its center. Each hemisphere corresponds to one half-space, with upper half considered as air and lower half as glass substrate. Far-field quantities radiated from the dipole are calculated considering NA 0.9 in both half-spaces. A spatial mesh resolution of 1 nm is employed to encompass the nanoantenna, with a refined mesh resolution of 0.5 nm applied in the nanogap region. Mesh convergence is evaluated by tracing the change in $IE_{R\text{-}CPL}$ and $IE_{L\text{-}CPL}$ spectra at the center of the nanogap under progressively refined nanogap mesh resolution ranging from 1 nm to 0.2 nm with the surrounding mesh fixed at 1 nm. The nanogap mesh resolution of 0.5 nm is adopted as a practical compromise between numerical accuracy and computational feasibility. Perfectly matched layer boundary conditions are used in all directions to eliminate spurious reflections. The simulation time and the built-in auto-shutoff criterion are chosen to ensure complete decay of the fields. All calculations in this work consider the monocrystalline gold permittivity from Olmon et al.[37] for the nanoantenna and the glass permittivity from Palik[38] for the substrate.

**Authors' Statements**

**Corresponding Authors**

**Abhik Chakraborty** − Institute of Physical Chemistry, Friedrich Schiller University Jena, Helmholtzweg 4, 07743 Jena, Germany; Abbe Center of Photonics, Friedrich Schiller University Jena, Albert-Einstein-Str. 6, 07745 Jena, Germany; Leibniz Institute of Photonic Technology, Albert-Einstein-Str. 9, 07745 Jena, Germany; E-mail: abhik.chakraborty@uni-jena.de; ORCID: https://orcid.org/0000-0001-6048-4201.

**Jer-Shing Huang** − Institute of Physical Chemistry, Friedrich Schiller University Jena, Helmholtzweg 4, 07743 Jena, Germany; Abbe Center of Photonics, Friedrich Schiller University Jena, Albert-Einstein-Str. 6, 07745 Jena, Germany; Leibniz Institute of Photonic Technology, Albert-Einstein-Str. 9, 07745 Jena, Germany; Research Center for Applied Sciences, Academia Sinica, 128 Sec. 2, Academia Road, Nankang District, Taipei 11529, Taiwan; Department of Electrophysics, National Yang Ming Chiao Tung University, Hsinchu 30010, Taiwan; E-mail: jer-shing.huang@leibniz-ipht.de; ORCID: https://orcid.org/0000-0002-7027-3042.

**Author Contributions**

A.C. conceived the idea, conducted numerical simulations, calculations and analysis, and wrote the manuscript. J.-S.H. supervised the project. All authors participated in discussions and contributed to manuscript revisions. All authors reviewed the results, approved the final version of the manuscript, and consented to the manuscript submission.

**Funding**

The work is funded by the Deutsche Forschungsgemeinschaft (DFG, German Research Foundation) via CRC/SFB 1375 NOA (project no. 398816777; sub-project C1) and IRTG 2675 "Meta-Active" (project no. 437527638; sub-project C1).

**Conflict of Interest**

The authors declare no conflict of interest.

**Data Availability Statement**

The data generated and analysed during the current study are available from the corresponding authors upon reasonable request.

# Supporting Information

## I. Modal Analysis of the Fully Symmetric, Partially Symmetric, and Planar Chiral Nanoantennas

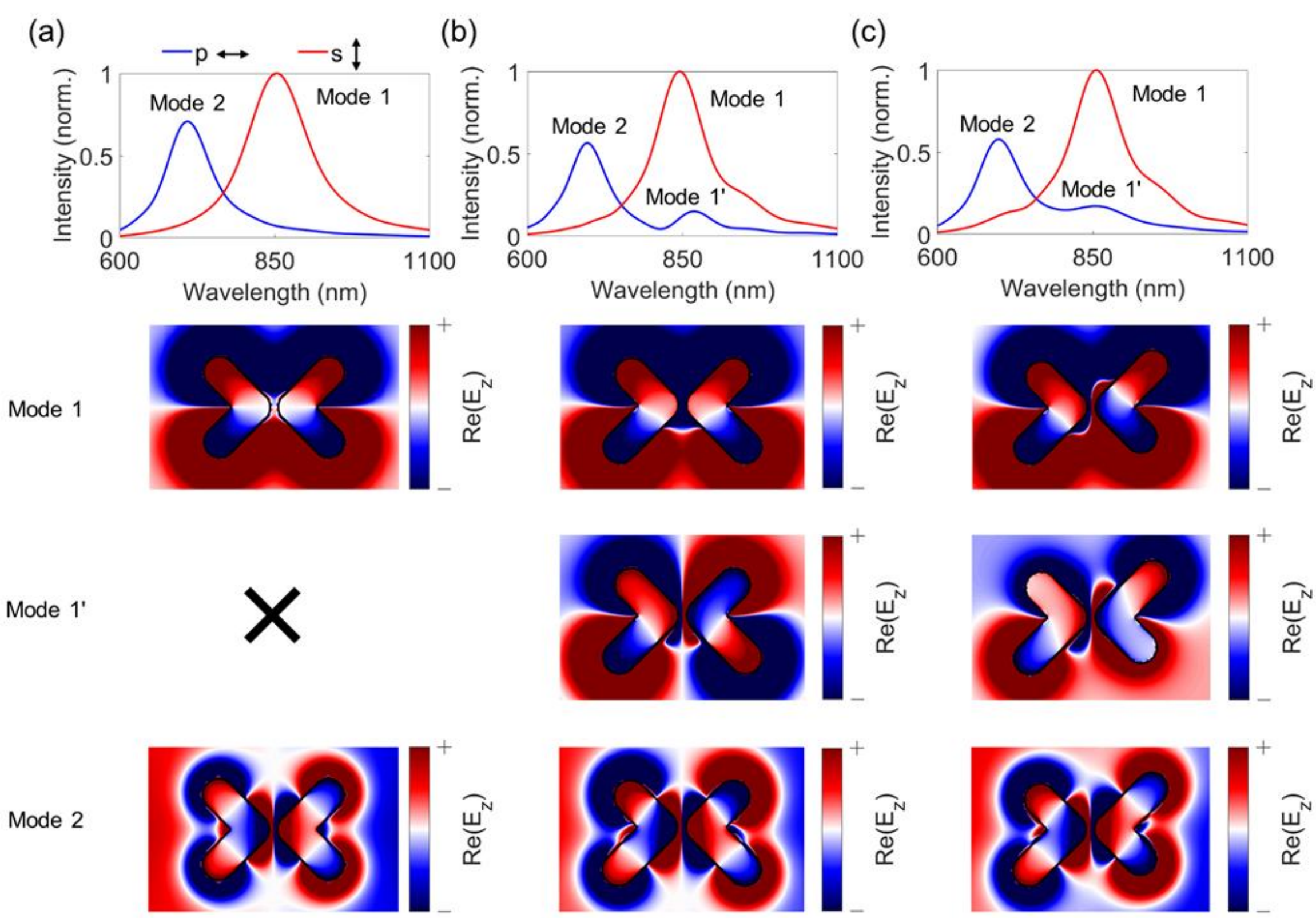

**Figure S1** Scattering spectra and modal field distributions ($Re(E_z)$) of the (a) fully symmetric (left column), (b) partially symmetric (middle column), and (c) planar chiral (right column) nanoantennas. The first row (top to bottom perspective) shows the normalized scattering spectra of (a) fully symmetric, (b) partially symmetric, and (c) planar chiral nanoantennas under p- (blue) and s-polarized (red) excitation. Mode 1 is the s-excited fundamental mode, Mode 1' is the p-excited fundamental mode, and Mode 2 is the p-excited higher-order mode. The second row exhibits the field distributions associated with Mode 1 in (a) fully symmetric, (b) partially symmetric, and (c) planar chiral nanoantennas. The third row exhibits the modal field distributions associated with Mode 1' in (b) partially symmetric and (c) planar chiral nanoantennas. The fourth row depicts the field distributions associated with Mode 2 in (a) fully symmetric, (b) partially symmetric, and (c) planar chiral nanoantennas.

To elucidate the origin of chirality, three nanoantenna geometries, i.e., fully symmetric (**Figure S1a**; left column of Figure S1), partially symmetric (Figure S1b; middle column), and planar chiral (Figure S1c; right column), are analyzed. The first row (top to bottom perspective) of Figure S1 depicts the normalized scattering spectra and dominant modes of the fully symmetric (Figure S1a), partially symmetric (Figure S1b), and planar chiral (Figure S1c) nanoantennas under p- (blue) and s-polarized (red) plane-wave excitation at normal incidence. p denotes the linear polarization of the excitation that is oriented along the axis of the nanoantenna's gap, whereas s denotes the excitation's linear polarization oriented orthogonally (in-plane) to the same axis. Here, the s-excited fundamental mode is marked as Mode 1, the p-excited fundamental mode is marked as Mode 1', and the p-excited higher-order mode is marked as Mode 2. The second row shows the field distributions ($\mathrm{Re}(\mathrm{E_z})$) associated with Mode 1 in fully symmetric (Figure S1a), partially symmetric (Figure S1b), and planar chiral (Figure S1c) nanoantennas. Similarly, the third row exhibits the field distributions ($\mathrm{Re}(\mathrm{E_z})$) associated with Mode 1' in partially symmetric (Figure S1b) and planar chiral (Figure S1c) nanoantennas. Finally, the fourth row exhibits the field distributions ($\mathrm{Re}(\mathrm{E_z})$) associated with Mode 2 in fully symmetric (Figure S1a), partially symmetric (Figure S1b), and planar chiral (Figure S1c) nanoantennas.

Even though $\mathrm{E_x}$ is the key to understanding excitation-chirality-controlled dissymetry in the x-axis-oriented nanogap due to the nanogap's geometry-induced confinement (**Figure 2** in the main manuscript), the real part of the out-of-plane field component ($\mathrm{Re}(\mathrm{E_z})$) is used to study the modal field distribution at a plane intersecting each planar nanoantenna 1 nm below its surface (top air−metal interface) in Figure S1. This is done because the out-of-plane component provides an unambiguous representation of the modal field distributions whereas in-plane components ($\mathrm{Re}(\mathrm{E_x})$ and $\mathrm{Re}(\mathrm{E_y})$) exhibit mixed polarization contributions in these complex planar geometries.

By comparing the analysis of the chiroptical response depicted in Figure 2 of the main manuscript with the spectral position of the modes shown in Figure S1, one can deduce that the simultaneously p- and s-excited fundamental modes, i.e., Mode 1' and Mode 1 respectively, with equal amplitude and appropriate phase difference co-polarized and co-localized in the nanogap determines the pronounced chiroptical response in our work. Figure S1 shows that symmetry-related constraints prohibit the emergence of Mode 1' in the fully symmetric nanoantenna (Figure S1a; third row). Furthermore, the fully symmetric nanoantenna geometry supports a symmetric field distribution corresponding to Mode 1 with nodes at the apexes of the two symmetric V-shaped elements, thereby producing minimal to no amplitude enhancement (AE) in the nanogap. Therefore, we can exclude the fully symmetric nanoantenna (Figure S1a) and Mode 2 (Figure S1; fourth row) from the discussion and concentrate the analysis exclusively on Mode 1 and Mode 1' in the partially symmetric nanoantenna and the planar chiral nanoantenna.

Anti-symmetric field distributions corresponding to Mode 1' are present in both partially symmetric (Figure S1b; third row) and planar chiral (Figure S1c; third row) nanoantennas. The anti-symmetric field distributions allow the emergence of opposite polarity at the apexes of the partially symmetric and planar chiral nanoantennas under p-polarized excitation, thereby facilitating x-polarized AE in both nanogaps (Figure 2b and Figure 2c in the main manuscript). Therefore, the missing piece in the analysis is understanding the role of structural asymmetry in endowing the s-excited Mode-1-induced AE in the nanogap.

The partially symmetric nanoantenna geometry (Figure S1b; second row) supports a symmetric field distribution corresponding to Mode 1 in the two asymmetric V-shaped elements. Due to the asymmetric geometry of each V-shaped element in the partially symmetric nanoantenna, the node is slightly displaced from the apex. However, the symmetric field distribution results in a weakly coupled anti-bonding mode (same field polarity at the apexes) and minimal x-polarized AE in the nanogap (Figure 2b in the main manuscript). The planar chiral nanoantenna geometry (Figure S1c; second row), due to its asymmetric V-shaped elements having a mirror-inverted geometrical configuration with respect to each other, supports an anti-symmetric field distribution corresponding to Mode 1 in the two V-shaped elements, thereby yielding opposite polarity at the apexes of the V-shaped elements. This enforces a bonding mode and strong x-polarized AE within the nanogap (Figure 2c in the main manuscript). The anti-symmetric field distribution of Mode 1 is identical to that of Mode 1', which is unique to the planar chiral nanoantenna.

The emergence of Mode-1- and Mode-1'-induced x-polarized AE in the planar chiral nanoantenna underlines the role of complete planar asymmetry in inducing spatially overlapping and co-polarized electric near fields under linearly polarized orthogonal excitations. This, along with optimally engineered relations between the two modes' x-polarized AE (Figure 2c and Figure 2e in the main manuscript) and phase difference (Figure 2d and Figure 2f in the main manuscript), facilitate the emergence of chiroptical response in the planar chiral nanoantenna.

## II. Radiated Intensity Spectra of Single Emitter-Driven Planar Chiral Nanoantenna

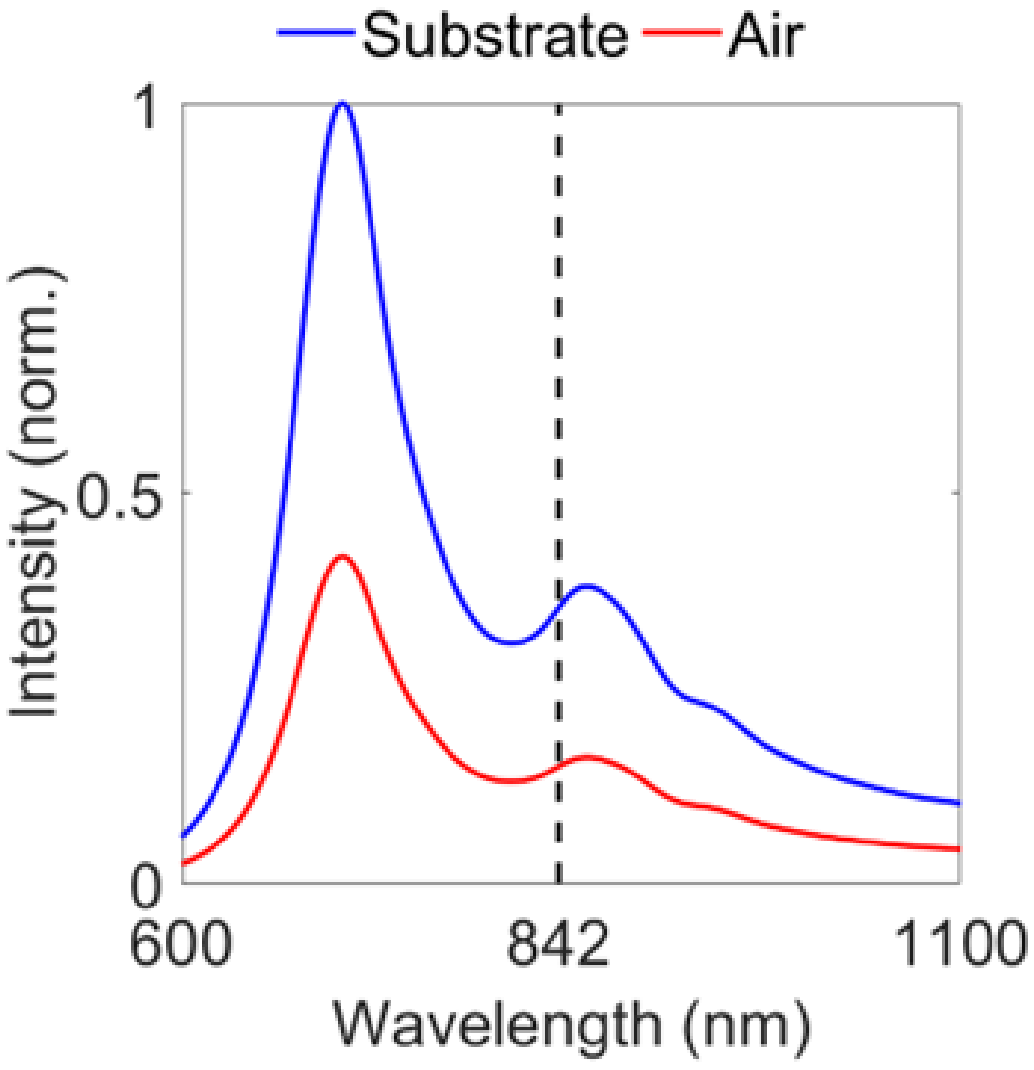

**Figure S2** Far-field intensity enhancement spectra radiated into substrate (blue) and air (red) directions from a dipole positioned at the geometric center of the planar chiral nanoantenna's nanogap and collected within a numerical aperture of 0.9.

Stokes parameters and, therefore, average degree of circular polarization (⟨DoCP⟩) are defined even in regions of space containing very low field. Therefore, while **Figure 4a** in the main manuscript exhibits very high ⟨DoCP⟩ values at and around 842 nm radiated into both the air side and the substrate side from an x-polarized dipole placed in the geometrical center of the planar chiral nanoantenna's nanogap, corresponding radiated intensity spectra are required to fully assess the efficiency of the dipole-in-nanoantenna system. **Figure S2** exhibits the normalized far-field intensity enhancement spectra of an x-polarized dipole placed in the nanoantenna's nanogap, with the nanoantenna situated on the substrate. The intensity radiated into a particular direction (substrate or air) is calculated within a numerical aperture (NA) of 0.9 and normalized by the intensity of the same dipole situated in free space radiated into the same direction and collected within the same NA. The spectra exhibited in Figure S2 are obtained by further normalizing the far-field intensity enhancement spectra by the highest value of radiated intensity recorded in the substrate. The wavelength of 842 nm (vertical dashed line) is marked by near-maximal ⟨DoCP⟩ in Figure 4a of the main manuscript, while also being very close to the fundamental modal peak of the nanoantenna's radiated intensity in Figure S2. This confirms the spectral region at and around 842 nm to be ideal for chiral emission from the optimally designed planar chiral nanoantenna. It is clear that the radiated intensity is higher into the substrate than that into air, with the intensity radiated into the substrate being 2.36 times that radiated into air at the wavelength of 842 nm.